\documentclass[conference]{IEEEtran}
\IEEEoverridecommandlockouts

\usepackage{cite}
\usepackage{amsmath,amssymb,amsfonts}
\usepackage{graphicx}
\usepackage{textcomp}
\usepackage[hyphens]{url}
\usepackage{fancyhdr}
\usepackage{hyperref}
\usepackage[normalem]{ulem}
\usepackage{pifont}
\usepackage{xspace}
\usepackage{xcolor}
\usepackage{soul}
\usepackage{enumitem}
\usepackage{lipsum}
\usepackage{textcomp}
\usepackage[shortcuts]{extdash}
\usepackage{booktabs}
\usepackage{multirow}
\usepackage{makecell}
\usepackage{algorithmic}
\usepackage{subfig}
\usepackage{tikz}
\usepackage{listings}

\newcommand{\fig}[1]{Figure~\ref{#1}}
\newcommand{\sect}[1]{Section~\ref{#1}}
\newcommand{\proposed}{SwarmIO\xspace}

\def\BibTeX{{\rm B\kern-.05em{\sc i\kern-.025em b}\kern-.08em
    T\kern-.1667em\lower.7ex\hbox{E}\kern-.125emX}}
\begin{document}

\title{\huge \proposed: Towards 100 Million IOPS SSD Emulation for \\Next-generation GPU-centric Storage Systems}

\author{\IEEEauthorblockN{Hyeseong Kim}
\IEEEauthorblockA{KAIST\\
\texttt{hyeseong.kim@kaist.ac.kr}}
\and
\IEEEauthorblockN{Gwangoo Yeo}
\IEEEauthorblockA{KAIST\\
\texttt{gwangoo525@kaist.ac.kr}}
\and
\IEEEauthorblockN{Minsoo Rhu}
\IEEEauthorblockA{KAIST\\
\texttt{mrhu@kaist.ac.kr}}
}

\maketitle

\begin{abstract}

GPU-initiated I/O has emerged as a key mechanism for achieving high-throughput storage access by leveraging massive GPU thread-level parallelism, while recent industry trends point toward SSDs optimized for ultra-high random-read IOPS. Together, these trends are enabling the emergence of IOPS-optimized, GPU-centric storage systems. Despite this momentum, no existing framework enables quantitative end-to-end evaluation of storage systems optimized for GPU-initiated I/O. While conventional SSD emulators provide a promising path toward end-to-end modeling in traditional storage systems, they face three key challenges in this GPU-centric setting: limited frontend scalability for ingesting massive request streams, high software overhead in emulating GPU-initiated I/O control and data paths, and excessive timing-model maintenance overhead at extremely high I/O request rates. We propose \proposed, an SSD emulator for massively parallel, GPU-centric storage. \proposed faithfully models IOPS-optimized SSDs at target performance levels of up to 40 MIOPS, achieving a 303.9$\times$ speedup over the state-of-the-art baseline SSD emulator under GPU-initiated I/O. We further demonstrate its utility through a vector search case study, showing that increasing SSD IOPS from 2.5 MIOPS to 40 MIOPS yields an average end-to-end speedup of up to 9.7$\times$.

\end{abstract}

\section{Introduction}
\label{sect:introduction}

The rapid growth of modern AI workloads has pushed runtime data footprints far beyond the capacity of GPU High Bandwidth Memory (HBM), making storage systems an essential component of the AI data pipeline. As applications evolve from simple generative queries to multi-stage, chain-of-thought agentic AI, the required Key-Value (KV) cache for massive context windows drives an unprecedented expansion in memory requirements. Given this landscape, storage is no longer merely a passive capacity layer, but is increasingly treated as an active, ephemeral memory tier for efficient large-scale AI deployments. To serve this role effectively, architectures such as the NVIDIA Inference Context Memory Storage Platform (CMX) have been introduced to act as an AI-native context tier, extending GPU memory across high-bandwidth networks~\cite{cmx}. This fundamental shift marks the collapse of the traditional memory-storage hierarchy, requiring NAND flash SSDs to operate at near-memory speeds to avoid computational stalling. 

Among the evolving performance demands on storage, an emerging class of data-intensive applications introduces a new design objective: sustaining high random read I/O operations per second (IOPS) for massively parallel, fine-grained accesses. Datacenter-native AI applications such as retrieval-augmented generation (RAG)~\cite{rag}, recommendation systems~\cite{dlrm}, and agentic AI workflows~\cite{hpca2026_agent} generate sparse, fine-grained accesses to large datasets. These workloads are driving the development of ``GPU-centric'' storage systems that prioritize fine-grained I/O parallelism, enabling GPUs to directly access storage via \emph{GPU-initiated I/O}~\cite{bam}. In this model, GPUs natively submit I/O requests on demand, bypassing the host CPU's orchestration path entirely. However, although GPU-initiated I/O can generate tens to hundreds of millions of IOPS (MIOPS), traditional storage devices cannot sustain such extreme random-read throughput.
For instance, even high-end enterprise-grade SSDs support only up to about 3 MIOPS~\cite{kioxia_cm9,micron_9550,solidigm_d7}, leaving current storage systems fundamentally unable to meet the massive, fine-grained IOPS demands of GPU-initiated I/O.

These limitations are motivating the development of next generation ``ultra-high IOPS'' storage architectures that prioritize IOPS over sequential bandwidth. Under initiatives such as NVIDIA’s StorageNext~\cite{storagenext}, industry leaders are rapidly developing IOPS-optimized SSD designs tailored for massively parallel, fine-grained 512-byte accesses from GPUs~\cite{storagenext,fms2025_nvidia,ocp2025_nvidia,sdc2025_nvidia,fms2025_kioxia,ocp2025_hynix,fms2025_fadu}. For example, the Kioxia GP Series, utilizing low-latency XL-FLASH~\cite{xlflash}, exemplifies this transition, targeting 10 MIOPS in 2026, with industry roadmaps projecting 100 MIOPS by 2027 using PCIe Gen6 and Gen7 interfaces~\cite{ocp2025_hynix,fms2025_kioxia}. Despite this momentum, system designers and application developers face a classic chicken-and-egg problem: while ultra-high IOPS storage systems based on next-generation SSDs are on the horizon, their practical benefits cannot yet be evaluated because the hardware is not commercially available. Nevertheless, quantifying the benefits of employing these ultra-high IOPS storage devices on end-to-end application performance is critical for guiding future system development, motivating the need for a quantitative, end-to-end performance modeling approach.

While cycle-level system simulators (e.g., gem5~\cite{gem5} integrated with GPGPU-Sim~\cite{gpgpusim,accelsim} and an SSD simulator~\cite{simplessd,mqsim}) could, in principle, provide such insights, they are prohibitively slow and cannot evaluate full-system behavior within a reasonable wall-clock time. For instance, GPGPU-Sim achieves a simulation speed of only about 3 KIPS (kilo-instructions per second), which is far too slow to simulate even a simple AI model's inference within a practical timeframe. A promising alternative is to develop a high-performance SSD \emph{emulator} capable of sustaining hundreds of millions of IOPS in ``real time'' while operating alongside a real GPU. Such an emulation framework enables end-to-end evaluation of future GPU-centric storage systems integrated with ultra-high IOPS SSDs. Unfortunately, existing SSD emulators~\cite{nvmevirt,femu,flexdrive,vssim} are inadequate for this regime as they face three key challenges. First, prior designs do not scale to tens or hundreds of millions of IOPS, as they were originally built for traditional CPU-based storage; their frontend architectures cannot efficiently ingest the massively parallel I/O request streams generated by GPUs, leading to severe queue buildup. Second, GPU-initiated I/O introduces distinct control and data paths that require software-mediated data movement between CPU-side emulated storage structures and GPU-resident I/O buffers, incurring significant overhead. Third, at such high request rates, per-request updates to the emulator’s timing model become a bottleneck, making it difficult to maintain high model fidelity at ultra-high IOPS.

\proposed is designed to fill this critical gap by providing an IOPS-scalable SSD emulation framework with a fundamental goal of reaching 100 MIOPS\footnote{\proposed is open-sourced at \href{https://github.com/VIA-Research/SwarmIO}{https://github.com/VIA-Research/SwarmIO.}}. \proposed is built upon three key innovations that significantly advance the performance of prior SSD emulators. First, we accelerate the frontend of our emulator using a parallelism-aware architecture that adopts a distributed software design and incorporates a throughput-oriented request fetching mechanism leveraging NVMe protocol semantics. This approach enables request ingestion throughput to scale with request-level parallelism while substantially reducing queuing delays. Second, we leverage the Intel Data Streaming Accelerator (DSA) to offload SSD backend data copy operations, significantly reducing the overhead associated with emulating GPU-initiated I/O. Our proposal is co-designed with a DSA-aware, kernel-level API for high-throughput copy offloading, maximizing DSA utilization in a multi-threaded environment. Third, we introduce an aggregated timing model update mechanism that amortizes state management overhead across a group of requests while preserving high-fidelity timing emulation.

We evaluate \proposed by first validating its modeling fidelity against an enterprise-grade 2.5 MIOPS SSD, and then demonstrating the scalability of its architecture. Although hardware constraints in our evaluation testbed cap the current emulation throughput at 40 MIOPS, this still represents a massive leap forward under extreme request parallelism unlocked with GPU-initiated I/O. We further show that \proposed enables end-to-end evaluation of AI applications that leverage GPU-initiated I/O, highlighting the need for GPU-centric storage systems capable of sustaining hundreds of millions of IOPS. Using vector search, a key component of RAG systems, as a case study, we quantify the end-to-end performance benefits of combining GPU-initiated I/O with ultra-high IOPS SSDs. We summarize the {\bf key contributions} of this work:

\begin{itemize}
  \item We propose \proposed, an IOPS-scalable NVMe SSD emulation framework that supports end-to-end modeling of GPU-centric storage systems with GPU-initiated I/O.

  \item We demonstrate, on real GPU systems, that \proposed scales to 40 MIOPS, achieving a 307.7$\times$ speedup over a state-of-the-art SSD emulation framework.

  \item We enable end-to-end analysis of GPU-centric storage systems with future IOPS-optimized SSDs, and demonstrate through a vector-search case study that increasing SSD IOPS from 2.5 MIOPS to 40 MIOPS yields an average 9.7$\times$ end-to-end system-level speedup.
\end{itemize}

\section{Background}
\label{sect:background}

\subsection{GPU-initiated Storage I/O}
\begin{figure}[t]
\centering
\includegraphics[width=0.49\textwidth]{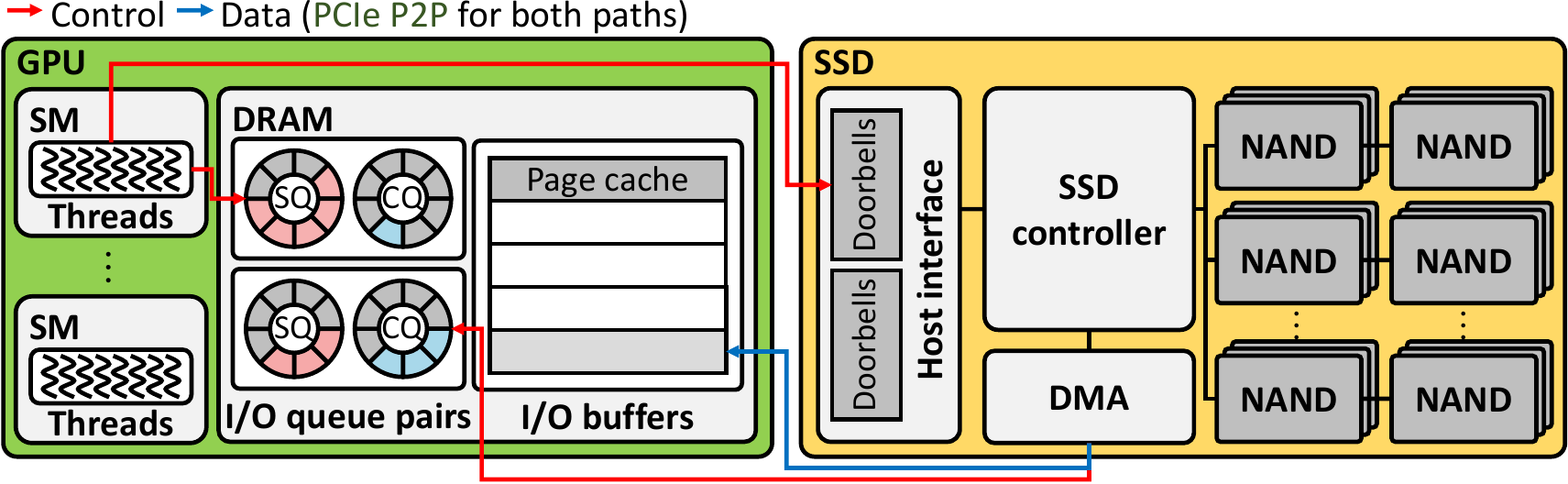} 
\caption{GPU-initiated I/O.}
\label{fig:gpu_initiated_io}
\end{figure}

GPU-initiated I/O enables high-throughput, fine-grained, on-demand storage accesses from the GPU, meeting the needs of emerging applications with sparse and random data access patterns~\cite{bam,gids,gmt}. As illustrated in \fig{fig:gpu_initiated_io}, GPU-initiated I/O offloads the entire I/O stack to the GPU, including I/O queues for the control path and I/O buffers for the data path. Consequently, all communication with the storage device occurs via PCIe peer-to-peer (P2P) transfers between the GPU and SSD, allowing the entire I/O lifecycle to proceed without CPU intervention. This design minimizes CPU--GPU synchronization overhead and reduces I/O amplification by enabling GPU threads to directly submit I/O requests on demand. By leveraging GPU's massive thread-level parallelism, GPU-initiated I/O can generate a large number of fine-grained I/O requests, fully stressing storage IOPS and PCIe bandwidth utilization beyond what traditional CPU-centric I/O can achieve for random accesses.

\subsection{SSD Simulation and Emulation Frameworks}

{\bf SSD simulators.} SSD simulators model storage I/O behavior, from the frontend host interface to the backend NAND flash device, within their own event time domains. Existing simulators generally fall into two categories. First, \emph{trace-driven} approaches~\cite{mqsim} replay timestamped I/O traces. Therefore, they lack closed-loop interaction with running systems, making it difficult to capture dynamic runtime behavior such as inter-request dependencies or interactions with host I/O stack components (e.g., the page cache). Second, \emph{full-system} approaches~\cite{mqsim,simplessd} integrate with full-system simulators~\cite{gem5} to enable detailed application-level characterization. 
However, these full-system simulators are not only slow but also ill-suited for studying GPU-initiated I/O; to the best of our knowledge, no existing full-system simulation framework models PCIe P2P between an SSD and a GPU.

{\bf SSD emulators.} In contrast, SSD emulators interact with running applications and can therefore provide end-to-end functional modeling, with several works~\cite{vssim,femu,flexdrive,nvmevirt} also supporting performance modeling in the wall-clock time domain. However, existing frameworks often fall short in scalability and interoperability with GPU-initiated I/O. Some emulators~\cite{flexdrive} require device-driver modifications to intercept I/O requests and emulate storage I/O using a DRAM-backed virtual disk while rate-limiting requests to match target SSD's performance. This approach is fundamentally incompatible with GPU-initiated I/O, which bypasses the host driver and uses a dedicated GPU-side I/O stack. Virtualization-based emulators~\cite{vssim,femu} instead present virtual SSDs to guest virtual machines via QEMU. However, they either incur significant context-switch overhead when trapping at MMIO operations~\cite{vssim} or lack support for the PCIe P2P transfers required by GPU-initiated I/O~\cite{femu}.

\subsection{NVMeVirt}
\label{subsect:nvmevirt}

Unlike prior SSD emulators, NVMeVirt~\cite{nvmevirt} exposes a software-defined PCIe device directly through the kernel PCI subsystem, presenting a \emph{virtual} SSD on a bare-metal host. This design enables the emulation of diverse storage environments, including those that require PCIe P2P transfers. As such, NVMeVirt is functionally capable of modeling GPU-initiated I/O. However, as detailed in \sect{sect:motivation}, it does not provide sufficient scalability to emulate future SSD IOPS targets, and the distinct control and data paths of GPU-initiated I/O further degrade achievable performance. We next describe the overall architecture and timing model of NVMeVirt.

\begin{figure}[t]
\centering
\includegraphics[width=0.48\textwidth]{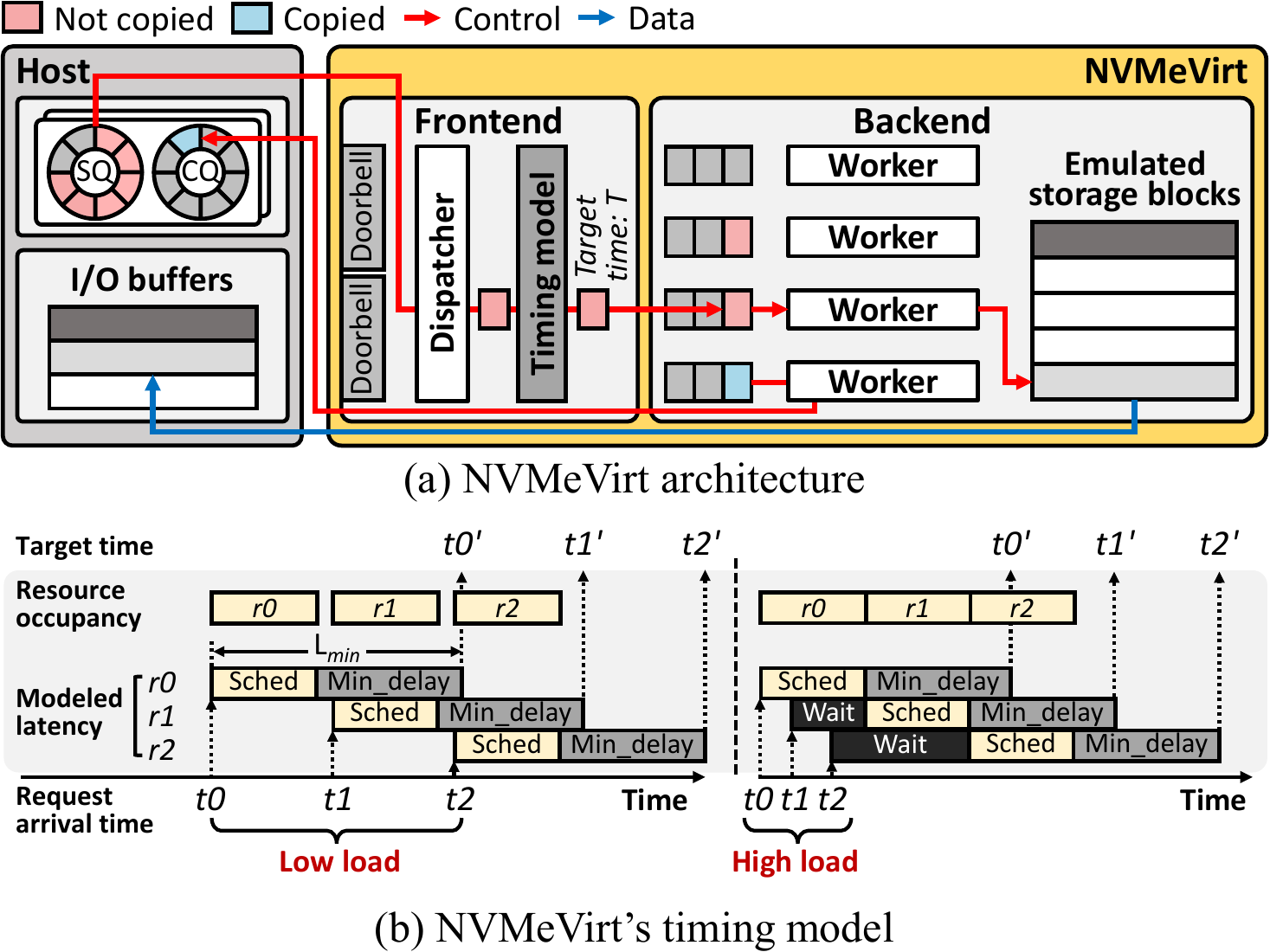}
\caption{(a) High-level overview of NVMeVirt and (b) an example illustrating its timing model. The timing model derives a request's target completion time when it is fetched by the dispatcher. \texttt{Sched} updates the modeled SSD resource availability to regulate sustained throughput (\texttt{T$_{max}$}), while \texttt{Min\_delay} enforces minimum per-request latency (\texttt{L$_{min}$}).}
\label{fig:nvmev_architecture}
\end{figure}

{\bf Overall architecture.} As shown in \fig{fig:nvmev_architecture}(a), NVMeVirt consists of a ``frontend'' that models the host interface and a ``backend'' that emulates storage data transfers, implemented by a single \emph{dispatcher} and multiple \emph{worker} threads, respectively. The dispatcher polls submission queue (SQ) doorbells (i.e., software-defined NVMe registers exposed to the host for signaling I/O submissions), and sequentially fetches newly submitted requests  when the doorbells are set. After fetching each request, it dispatches the request to one of the workers via its per-worker queue, where storage data transfers are emulated by deriving the target completion time of the request using NVMeVirt’s \emph{timing model}.
A worker scans all its local queue entries in each iteration and checks the status of each request. If a request is newly enqueued, the worker emulates the corresponding storage data transfer by copying data between the target I/O buffer and the emulated storage blocks, which is implemented by modeling the storage block address space in CPU memory. A worker posts completion of a request to the  completion queue (CQ) only if the target completion time has already elapsed. Otherwise, the request remains copied but not yet completed. For such requests, the worker re-evaluates whether the target completion time has been elapsed in subsequent iterations.

{\bf Timing model.} Among the timing models provided by NVMeVirt, we describe its simple timing model as a representative example. NVMeVirt captures SSD performance using two configurable parameters: maximum throughput (\texttt{T$_{max}$}) and minimum latency (\texttt{L$_{min}$}). 
It abstracts SSD hardware resources as a set of parallel scheduling instances (e.g., flash controllers and channels) and decomposes each I/O request into unit-sized operations (e.g., read from flash page) that are scheduled across these scheduling instances. By tracking scheduling instance availability (i.e., resource occupancy), the timing model regulates sustained throughput below the configured maximum throughput (\texttt{T$_{max}$}), and each request is assumed to occupy its target instances for a fixed scheduling time (\texttt{Sched}), during which the instance is unavailable to others. \fig{fig:nvmev_architecture}(b) illustrates this process assuming a single scheduling instance. Under high input load, if the target instance is occupied by preceding requests when a request arrives, scheduling is deferred until the instance becomes available; for example, the instance is occupied by request $r0$ when $r2$ arrives at $t2$. The target completion time is then derived by adding \texttt{Sched} and an additional delay (\texttt{Min\_delay}) to enforce the configured minimum latency (\texttt{L$_{min}$}). Under low input load, by contrast, a request is scheduled as soon as it arrives, and its target latency is simply the sum of the two delays, \texttt{Sched} and \texttt{Min\_delay}.

\subsection{Intel Data Streaming Accelerator}
\label{subsect:dsa_architecture}
The Intel Data Streaming Accelerator (DSA)~\cite{dsa}, integrated into 4th-Gen Xeon Scalable Processors, accelerates data movement and transformation tasks. DSA is organized around \emph{groups}, which are software-configurable units comprising \emph{work queues} (WQs) and \emph{engines} (blue box in \fig{fig:swarmio_overview}). Software offloads operations by issuing 64-byte work \emph{descriptors} to specific WQs, with each descriptor occupying one WQ slot. Descriptors are issued via 64-byte writes to portals, which are MMIO regions mapped to specific WQs. WQs can operate in either shared mode, where multiple software clients share a queue, or dedicated mode, where a single client exclusively uses the queue and manages available descriptor slots itself. Engines fetch descriptors from any WQ within the same group and execute the corresponding operations in a pipelined manner. Among the supported descriptor types, our work uses memory copy and batch descriptors to offload copy operations during emulation, where batch descriptors allow an array of copy descriptors to be issued at once, amortizing descriptor issue overhead~\cite{2024_asplos_dsa}. Each copy descriptor can use either physical addresses (PAs) and virtual addresses (VAs) to specify the source and destination locations.

\section{Characterization and Motivation}
\label{sect:motivation}

In this section, we characterize the limitations of existing SSD emulators in modeling future IOPS-optimized SSDs while supporting GPU-initiated I/O. We focus on NVMeVirt, as alternative emulators are incompatible with GPU-initiated I/O due to their reliance on host-system modifications or virtualization mechanisms. As we detail in this section, although NVMeVirt provides the functional foundation for modeling GPU-initiated I/O by presenting a software-defined, virtual PCIe device to the host, it cannot sustain the level of request-level parallelism required by future SSDs.

\subsection{Frontend Scalability Bottleneck}
\label{subsect:fe_scalability}

Modern SSDs adopt NVMe as the storage protocol because it is designed to fully exploit SSD throughput via a multi-queue I/O interface. An SSD controller includes a host interface layer (i.e., the frontend) that dispatches incoming requests and handles completions through I/O queue pairs. To fully leverage NVMe’s request-level parallelism within and across I/O queues, the frontend request ingestion bandwidth must not become a performance bottleneck. In practice, several works~\cite{openssd,marvell,openexpress,nvmecha} accelerate this layer using dedicated hardware to mitigate this bottleneck.

\begin{figure}[t]
\centering
\includegraphics[width=0.42\textwidth]{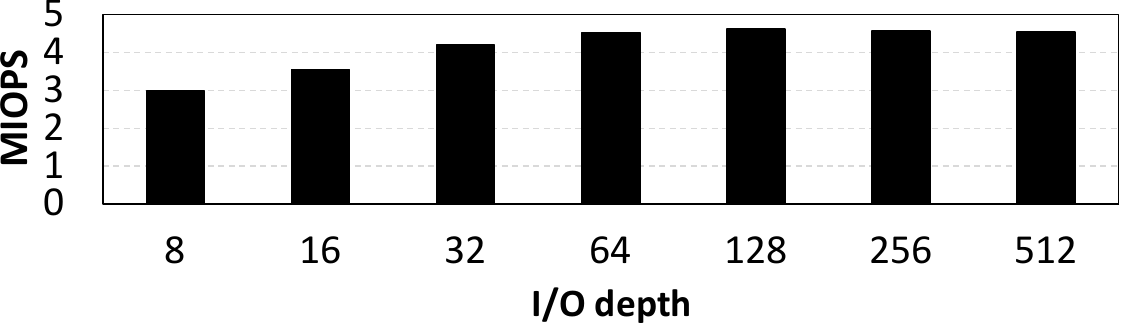} 
\caption{Frontend throughput of NVMeVirt under CPU-centric I/O with 32 fio threads.
}
\label{fig:nvmev_fio_dispatcher_iops}
\end{figure}

{\bf Limited dispatch throughput.} Despite its importance, frontend performance scalability is often overlooked in existing software-based SSD emulators, which struggle to sustain highly concurrent I/O requests arriving across different SQs. In particular, NVMeVirt relies on a \emph{single} dispatcher thread that sequentially polls the emulated doorbells of all SQs, which is similar to frontend designs commonly adopted in prior work~\cite{femu}. When multiple SQ doorbells are updated simultaneously, the dispatcher  fetches requests from a given SQ before proceeding to the next, thereby serializing I/O request processing. Consequently, the centralized frontend architecture exacerbates SQ queuing delays and prevents emulation throughput from scaling with the request-level parallelism exposed by the NVMe interface.

We use the fio~\cite{fio} 4 KB random read benchmark to evaluate NVMeVirt’s scalability under traditional ``CPU-centric'' I/O. The dispatcher and worker threads are pinned to dedicated cores within a single CPU socket, while fio runs on the remaining cores in the same socket. \sect{sect:methodology} provides further details of our experimental setup. To isolate frontend dispatcher performance, we disable worker-side data transfer emulation, which we refer to as the backend (\fig{fig:nvmev_architecture}(a)). We then stress the dispatcher by running 32 fio threads in parallel, each submitting I/O requests through its own dedicated SQ, while increasing the I/O depth from 8 to 512. A higher I/O depth denotes a larger number of outstanding requests per thread within its SQ, thereby increasing the submission pressure on NVMeVirt. As shown in \fig{fig:nvmev_fio_dispatcher_iops}, the frontend dispatcher’s throughput plateaus at 4.6 MIOPS despite increasing I/O depth, well below the 10--100 MIOPS target range of future ultra-high IOPS SSDs. \emph{This limited frontend scalability fundamentally constrains achievable throughput to only a few MIOPS, regardless of how effectively backend data transfers are parallelized across multiple workers.}

\subsection{Key Challenges with GPU-initiated I/O}
\label{subsect:challenges_with_bam}

Beyond modeling ultra-high IOPS SSDs, this work aims to provide an end-to-end emulation framework for GPU-initiated I/O. However, GPU-initiated I/O's control and data paths differ fundamentally from those of traditional CPU-centric I/O, posing unique challenges for faithfully modeling this emerging storage paradigm for real time emulation.

{\bf Small-block data transfers over PCIe.} As described in \sect{subsect:nvmevirt}, NVMeVirt emulates storage data transfers by having worker threads copy data between the target I/O buffers and emulated storage blocks. Under traditional CPU-centric I/O, both ends of this copy reside in CPU memory. Under GPU-initiated I/O, however, both the I/O buffers and I/O queue pairs (SQ and CQ) reside in GPU memory (\fig{fig:gpu_initiated_io}), so each request requires data transfer over PCIe, introducing additional overhead in both the data and control paths. In the data path, applications using GPU-initiated I/O often exhibit fine-grained access granularity, typically below 8~KB~\cite{fms2025_nvidia}, forcing workers to perform many small PCIe transfers (e.g., CPU-to-GPU copies for storage reads). In the control path, the dispatcher sequentially fetches even smaller 64-byte SQ entries allocated in GPU memory, resulting in fragmented PCIe transactions. \emph{Therefore, when combined with limited CPU thread-level parallelism, NVMeVirt fails to effectively utilize PCIe bandwidth, limiting achievable IOPS.}

{\bf Address mapping overhead.} 
Natively, GPU-initiated I/O allows SSDs to access GPU memory by exposing it in the host physical address (PA) space through a base address register (BAR)~\cite{gpu_direct_rdma}. A GPU thread submits an NVMe command that specifies the target I/O buffer with a PA, and the SSD's DMA engine uses that PA to perform P2P DMA to and from GPU memory. In NVMeVirt, however, data transfers are emulated in software using CPU worker threads. Upon receiving a request from the dispatcher, an NVMeVirt worker must map the target I/O buffer’s PA to a virtual address (VA) to access the GPU memory from the CPU. Because the target I/O buffer address differs across requests, workers must dynamically map and unmap it for each emulated data transfer using \texttt{memremap()} and \texttt{memunmap()}, incurring substantial latency overhead.

\begin{figure}[t]
\centering
\includegraphics[width=0.42\textwidth]{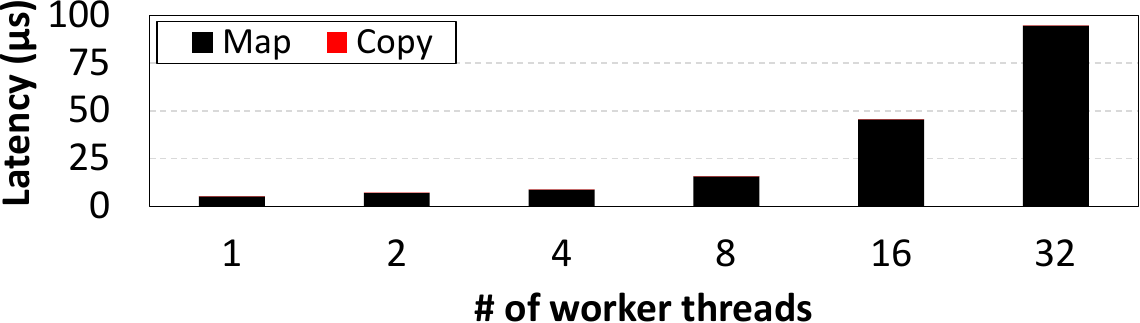}
\caption{Dynamic address mapping/unmapping overhead (\texttt{Map}) to prepare CPU thread-driven 512-byte random data copies (\texttt{Copy}) from CPU to GPU memory.}
\label{fig:nvmev_map_overhead}
\end{figure}

Because these functions modify page tables under internal kernel locks, \emph{this repeated mapping and unmapping by numerous worker threads creates a severe serialization bottleneck.} To quantify the resulting overhead, we implement a microbenchmark that models NVMeVirt’s backend data transfer behavior under GPU-initiated I/O while assuming the frontend is not a performance bottleneck (i.e., a worker-only setup). Multiple CPU worker threads repeatedly map a target buffer, copy a random 512-byte block from the CPU-side emulated storage blocks to the GPU-side I/O buffer, and unmap the buffer. As shown in \fig{fig:nvmev_map_overhead}, mapping and unmapping (\texttt{Map}) account for 98.8\% of the total data transfer latency on average. Under heavy contention with 32 threads, the latency of emulating a single storage data transfer reaches 94 $\mu\text{s}$, limiting aggregate throughput to only 0.34 MIOPS (i.e., at most 32 parallel I/Os per 94 $\mu\text{s}$), far below the 3 MIOPS available in state-of-the-art SSDs~\cite{kioxia_cm9,micron_9550,solidigm_d7}.

\begin{figure}[t]
\centering
\includegraphics[width=0.48\textwidth]{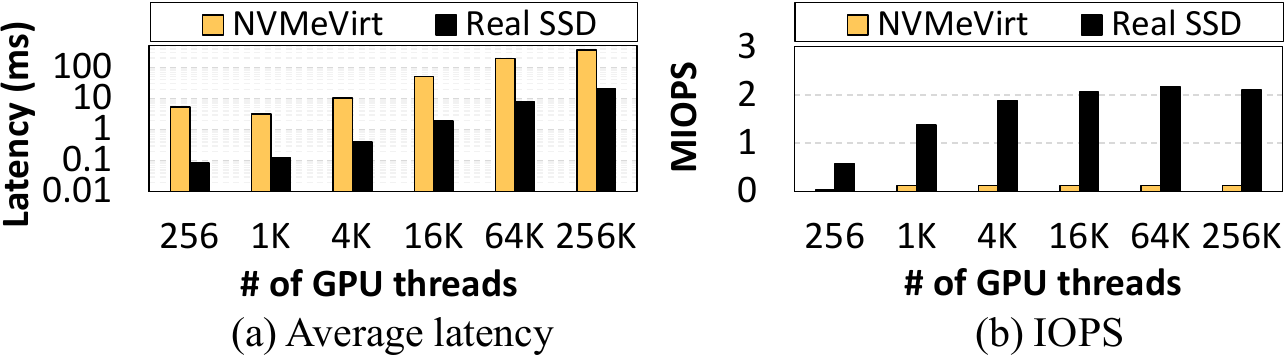}
\caption{(a) Average latency and (b) IOPS of NVMeVirt under GPU-initiated I/O, compared with those observed under a real SSD (Solidigm D7-PS1010), as the number of GPU threads submitting 512-byte random read I/O requests varies.}
\label{fig:nvmev_bam_performance}
\end{figure}

{\bf NVMeVirt vs. real SSD.} Using the BaM~\cite{bam} 512-byte random read benchmark on an NVIDIA H200 GPU~\cite{h200}, we evaluate NVMeVirt’s end-to-end performance under GPU-initiated I/O while varying the number of GPU threads submitting storage read I/Os. NVMeVirt’s timing model is configured to match that of a real Solidigm D7-PS1010 SSD. As shown in \fig{fig:nvmev_bam_performance}, NVMeVirt achieves at most 0.13 MIOPS with 256K GPU threads, making it 16.6$\times$ slower than the same benchmark running on the real SSD. \emph{Overall, NVMeVirt incurs 17.4--63.7$\times$ higher latency than the physical SSD, failing to meet our real-time emulation requirements for next-generation GPU-centric storage systems.}

\section{\proposed}
\label{sect:proposed}

In this section, we present \proposed, an IOPS-scalable SSD emulation framework for GPU-centric storage systems. Existing SSD emulators fail to faithfully model end-to-end performance of GPU-centric storage: some lack support for GPU-initiated I/O, while others (e.g., NVMeVirt) provide only functional modeling and suffer from non-scalable frontends and high CPU-driven data movement overheads. \proposed addresses these limitations with a distributed, high-throughput frontend design, hardware-accelerated data transfers, and a new timing model update mechanism that preserves emulation fidelity at extreme IOPS. Together, these techniques remove fundamental bottlenecks in prior emulators, significantly improving achievable IOPS for GPU-initiated I/O.

\subsection{High-level Overview}
\label{subsect:overview}

\begin{figure}[t]
\centering
\includegraphics[width=0.46\textwidth]{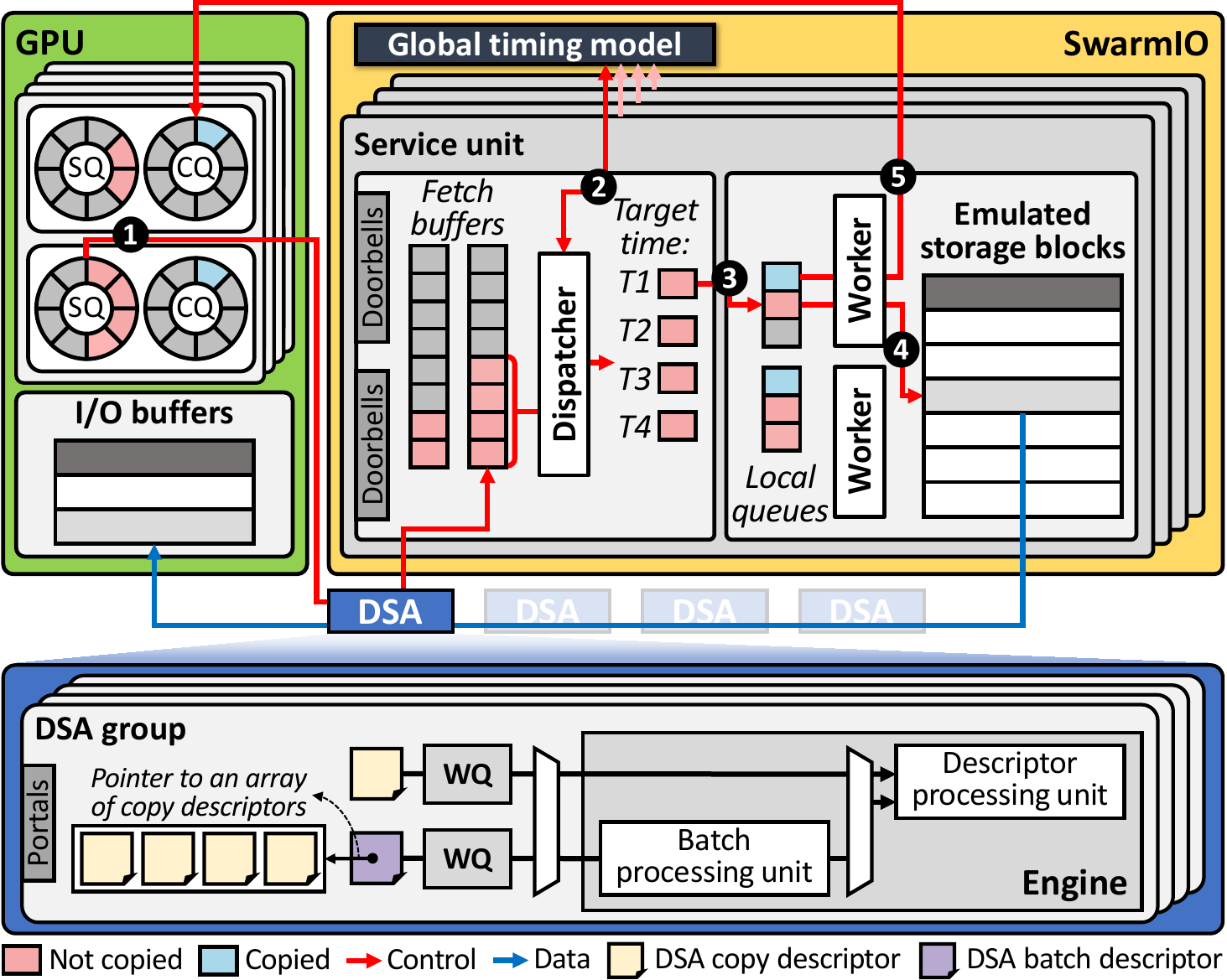}
\caption{High-level overview of \proposed.}
\label{fig:swarmio_overview}
\end{figure}

\proposed is implemented as a Linux kernel module and builds on NVMeVirt’s core mechanisms to provide a virtual SSD abstraction within the PCI subsystem. Prior work~\cite{femu,nvmevirt} relies on centralized frontends and throughput-limited data transfer emulation. In contrast, \proposed introduces a distributed, IOPS-oriented architecture centered around a modular operational unit called the \emph{service unit} (\fig{fig:swarmio_overview}).

Each service unit adopts a producer--consumer model where a dispatcher (frontend) polls SQ doorbells to fetch requests, while multiple workers (backend) process storage data transfers in parallel and handle completions. \fig{fig:swarmio_overview} depicts the execution flow within a service unit: the dispatcher first fetches enqueued requests from its assigned SQs (step \ding{182}) and computes their target completion times by invoking a global timing model shared across all service units (step \ding{183}). These timestamped requests (\texttt{T1-4}) are then distributed to the local queues of assigned workers (step \ding{184}). Each worker operates in a continuous loop: it first performs storage data transfers for up to a predefined number of requests from its local queue (step \ding{185}), and then re-evaluates the queue from the beginning to process completions for any ``copy-done'' requests whose target completion times have arrived (step \ding{186}). This mechanism ensures that request completion is not delayed by a long sequence of preceding data transfers, thereby maintaining strict adherence to the timing model.

By partitioning I/O queue pairs across service units, we transform the centralized frontend into a distributed architecture. With coalesced request fetching, \proposed fully exploits NVMe request-level parallelism (\sect{subsect:parallelism_aware_frontend}). In addition, to reduce the software overhead of CPU-driven data copies, \proposed adopts DSA-accelerated storage I/O emulation (\sect{subsect:dsa_accelerated_data_transfer}). It offloads both dispatcher\-/side request fetching and worker-side storage data transfers to Intel DSAs~\cite{dsa}, and introduces a DSA-aware kernel-level API for high-throughput copy offloading in multi-threaded environments. Finally, while the global timing model shared across dispatchers enables consistent performance modeling, it also introduces a nontrivial design challenge: under high request rates, frequent updates to the shared timing state by all dispatchers can incur substantial serialization overhead. To address this challenge, we introduce aggregated timing model updates (\sect{subsect:aggregated_timing_model_updates}), which minimize serialization across dispatchers by entering the critical section only once per set of requests. Together, these techniques enable \proposed to push achievable IOPS far beyond the limits of prior SSD emulators under massive request-level parallelism in GPU-centric storage systems.

\subsection{Parallelism-aware Frontend Architecture}
\label{subsect:parallelism_aware_frontend}

{\bf Distributed dispatch architecture.} To overcome frontend bottlenecks, \proposed partitions NVMe I/O queue pairs (SQ/CQ) across multiple service units, each led by a dedicated dispatcher thread (\fig{fig:swarmio_overview}), reducing the number of SQs handled by each dispatcher. This alleviates SQ queuing delay when a single dispatcher cannot keep up with requests submitted to tens to hundreds of SQs by multiple host (or GPU) threads. Each service unit is provisioned with dedicated compute resources, including worker threads and DSAs, as detailed in \sect{subsect:dsa_accelerated_data_transfer}. This design ensures consistent performance across I/O request flows targeting different sets of I/O queues and scales flexibly by instantiating the optimal number of service units proportional to the target storage IOPS.

{\bf Coalesced request fetching.} Beyond inter-queue parallelism from distributed dispatching, \proposed exploits \emph{intra}-queue parallelism via coalesced request fetching. Upon an SQ doorbell update, a dispatcher identifies newly enqueued entries and fetches multiple requests over a \emph{single} transfer into a reserved CPU memory region, the fetch buffer (\fig{fig:swarmio_overview}). This optimization is particularly effective under GPU-initiated I/O for two reasons. First, coalesced request fetching naturally aligns with the submission semantics of GPU-initiated I/O where requests within a warp are submitted together to the same SQ via a single doorbell update~\cite{bam}. Meanwhile, many warps interleave their accesses to SQ, causing a large number of requests to accumulate in each SQ and creating ample opportunity to fetch these requests at once. Second, because request fetching in GPU-initiated I/O requires GPU-to-CPU data transfer, coalesced fetching can merge multiple small PCIe transactions, provided that SQ entries are physically contiguous. To enable this, \proposed sets the Contiguous Queues Required (CQR)~\cite{nvme} bit to ‘1’ in the emulated controller register, as defined by the NVMe specification, ensuring that all entries within each SQ/CQ are allocated contiguously in physical memory. By merging multiple PCIe transactions, \proposed amortizes transaction overhead and improves transfer efficiency. Together with distributed dispatching, coalesced fetching enables \proposed to efficiently handle bursty I/O submissions while maintaining a highly parallel and scalable frontend design.

\subsection{DSA-accelerated Storage I/O Emulation}
\label{subsect:dsa_accelerated_data_transfer}

\subsubsection{Design Principles for DSA-efficient Data Movement}
\label{subsubsect:dsa_accelerated_design}
\mbox{}\\
\noindent{\bf Rethinking data transfers with DSA.} Despite improved frontend scalability for parallel request ingestion, software-mediated data transfers over PCIe in GPU-initiated I/O remain a major performance bottleneck, as discussed in \sect{subsect:challenges_with_bam}. To address this, we propose hardware-accelerated data movement using Intel DSA, which offloads data transfers from the CPU and enables total IOPS to scale beyond CPU processing limits. Importantly, offloading worker-side copies to DSA eliminates address mapping overhead, as DSA operates directly on PAs without requiring VA mappings. While NVMeVirt utilizes the legacy Crystal Beach DMA (CBDMA) engine to accelerate data transfers, its support is limited to na\"ively offloading worker-side transfers via a synchronous issue-and-wait model. We confirmed that this approach provides minimal practical benefit in reducing PCIe transfer latency. Moreover, prior studies~\cite{2024_isca_intel_acc,2024_asplos_dsa} show that even with DSA, such synchronous, per-request offloading fails to amortize the software overheads in preparing and issuing individual copy descriptors. This limitation is more severe for CBDMA, which incurs higher offloading overheads than DSA.

\proposed is carefully designed to overcome these limitations by enabling highly parallel data transfers that exploit key architectural features of DSA (\sect{subsect:dsa_architecture}). First, workers use \emph{descriptor batching} and \emph{asynchronous offloading} to maximize DSA utilization. Each worker receives requests from the dispatcher via its local queue. Rather than issuing copy descriptors for each request, it collects multiple requests, up to a predefined maximum batch size, into a single DSA \emph{batch descriptor} (purple in \fig{fig:swarmio_overview}) to amortize the overhead of issuing copy descriptors. In addition, \proposed maintains multiple in-flight batch descriptors via asynchronous offloading to overlap CPU-thread-side computation with DSA-side data movement, while exploiting the DSA engine's pipelined execution to overlap data transfers across batch descriptors. Second, \proposed extends DSA acceleration not only to worker-side data transfer emulation but also to dispatcher-side request fetching. Combined with coalesced request fetching, this creates a significant opportunity to improve transfer efficiency by enabling each P2P DMA transaction to fetch up to tens of KB. Together, these techniques enable \proposed{}'s frontend to sustain tens of MIOPS, which would not be achievable without either technique (see \sect{subsect:ablation_study} for an ablation study). Overall, our DSA-accelerated data transfers effectively mitigate software overhead and establish high-throughput control and data paths for storage I/O emulation.

{\bf DSA-aware kernel-level API for copy offloading.} To fully unlock DSA's potential, \proposed requires a high-performance kernel-level programming interface for offloading data transfers because it is implemented as a kernel module. The standard Linux DMA Engine API~\cite{dmaengine} exposes diverse DMA controllers via ``hardware-agnostic'' abstractions. While this abstraction simplifies DMA usage, it fails to fully exploit DSA's capabilities for three key reasons. First, the DMA Engine API does not expose DSA-specific features. For example, although the DSA device driver (\texttt{idxd}) provides low-level device control, the DMA Engine API exposes little beyond \texttt{idxd\_dma\_submit\_memcpy()}, which supports only \emph{non-batched} offloading of a single copy. Second, the default API solely relies on interrupt-driven notifications, even though DSA also supports polling. This interrupt-based design hinders latency-sensitive completion handling, even though \proposed requires timely request completions to preserve timing model fidelity. While polling preserves low latency, it incurs additional CPU overhead; however, this cost can be effectively hidden by issuing requests up to DSA’s maximum concurrency, overlapping them with useful computation, and polling only when necessary. Finally, because the API does not maintain per-thread offloading context (e.g., preallocated descriptor arrays and per-descriptor status), each thread must repeatedly allocate and program DSA descriptors for every request. This incurs unnecessary control path overhead and prevents full utilization of the multi-threaded execution environment.

\begin{figure}[t]
\centering
\begin{lstlisting}[
    language=C,
    basicstyle=\ttfamily\fontsize{6.4}{7.8}\selectfont,
    keywordstyle=\color{blue},
    stringstyle=\color{red},
    commentstyle=\color{green!60!black},
    emphstyle=\color{magenta},
    aboveskip=0pt,
    belowskip=0pt,
    frame=single,
    breaklines=true,
    captionpos=b,
    keepspaces=true,
    columns=fullflexible,
    numbers=left,
    numbersep=5pt,
    numberstyle=\tiny\color{gray},
    tabsize=2,
    xleftmargin=9pt,
    xrightmargin=7.5pt,
    emph={
        do_async_batch_dsa_offloading,
        empty, head, do_something,
        dsa_ctx_init,dsa_batch_issue_async,
        dsa_batch_should_issue_pending,
        dsa_batch_issue_pending,
        dsa_batch_should_wait,
        dsa_batch_wait_oldest},
    morekeywords={bool}
]
void do_async_batch_dsa_offloading(void) {
  ...
  struct dsa_ctx *ctx; /* Setup per-thread DMA context */
  dsa_ctx_init(ctx, dev, wq, batch_size, num_desc, ...);
    
  while (!empty(queue)) {
    for (i = 0; i < n; i++) { /* Add requests into a batch */
      work = head(queue);
      dsa_batch_issue_async(ctx, work.dest, work.src, 
                            work.len, ...);
    }
    do_something(); /* Perform useful computation */

    /* Issue the pending batch on timeout */
    if (dsa_batch_should_issue_pending(ctx, s_timeout))
        dsa_batch_issue_pending(ctx);
        
    /* Wait for the oldest in-flight batch on timeout */  
    if (dsa_batch_should_wait(ctx, c_timeout))
        dsa_batch_wait_oldest(ctx);
  }
}
\end{lstlisting}
\caption{An example asynchronous, batched data transfer workflow using our DSA-aware kernel-level offloading API.}
\label{fig:dma_batch_async_api}
\end{figure}

To address these challenges, we design a DSA-aware kernel-level API that streamlines copy offloading in multi-threaded environments. As shown in \fig{fig:dma_batch_async_api}, the API operates via a per-thread offloading context (i.e., \texttt{struct dsa\_ctx} in line 3). During initialization (\texttt{dsa\_ctx\_init} in line 4), each thread configures its dedicated context (\texttt{ctx}) by specifying (1) the target DSA WQ (\texttt{wq}), (2) the maximum batch size (\texttt{batch\_size}), and (3) the descriptor count (\texttt{num\_desc}), denoting the per-thread WQ depth (i.e., the maximum number of in-flight descriptors maintained in the WQ) for asynchronous offloading. The API also preallocates descriptors for reuse across requests and provides interfaces for directly programming DSA descriptors, including their use as batch descriptors. A representative asynchronous workflow enabled by our API is as follows:

\begin{enumerate}
\item{\textit{Batch descriptor construction} (line 7--11): A thread collects multiple copy operations into a single batch descriptor, and the API transparently issues the batch descriptor once the maximum batch size is reached (i.e., \texttt{i} $\ge$ \texttt{batch\_size}).}

\item{\textit{Computation pipelining} (line 12): Instead of busy-waiting, the thread performs useful computation while DSA processes data transfers. For instance, after issuing new batch descriptors, \proposed workers re-evaluate their local queues to process requests whose copies have completed and target completion times have been reached.}

\item{\textit{Timeout-driven issue} (line 15--16): To ensure forward progress under low load, the thread explicitly issues the current batch when pending entries remain and the configured timeout expires, as tracked by the DSA context.}

\item{\textit{Timeout-driven waiting} (line 19--20): Once sufficient time has elapsed, the thread polls for completion of the oldest in-flight batch, thereby reducing unnecessary blocking.}
\end{enumerate}

Through this concurrent workflow, \proposed amortizes the overhead of issuing copy descriptors and overlaps useful computation with DSA-orchestrated data transfers, thereby overcoming the limitations of baseline NVMeVirt, which primarily relies on CPU thread-level parallelism for data transfers.

\subsubsection{DSA Configuration Tuning via Microbenchmarking}
\label{subsubsect:ubenchmark}
\mbox{}\\
\noindent {\bf Microbenchmark.}  To validate our API design and identify optimal DSA configurations for \proposed, this subsection characterizes DSA performance using a microbenchmark that models GPU-initiated I/O. At a high level, our microbenchmark fetches groups of I/O requests from the SQ (GPU$\rightarrow$CPU), followed by the corresponding storage I/O operations (CPU$\rightarrow$GPU), with the goal of identifying the design points that maximize the throughput of frontend and backend of \proposed. The microbenchmark utilizes our service unit abstraction that mirrors \proposed{}'s design. In each service unit, the (frontend) dispatcher  performs sequential 4 KB GPU$\rightarrow$CPU transfers to model coalesced fetching of 64 requests, while the (backend) workers perform strided 512-byte CPU$\rightarrow$GPU transfers to model random 512-byte storage reads. Each service unit is mapped to a DSA group with one engine and two WQs, as shown in \fig{fig:swarmio_overview}. Our microbenchmark instantiates four service units over a single DSA device to fully exploit the maximum of four DSA groups supported by our device. Although the dispatcher and workers operate independently in this microbenchmark, our goal is to identify the operating point that balances frontend and backend IOPS while maximizing overall system-level IOPS. Because one 4 KB dispatcher transfer drives 64 worker-side transfers in \proposed, we focus on maximizing worker IOPS; accordingly, we use asynchronous, batched offloading to DSA for workers and synchronous, non-batched offloading for the dispatcher. We limit total dispatcher IOPS to 20 MIOPS (counting each 4 KB transfer as 64 I/O requests) and seek to maximize total worker IOPS up to this threshold. 

\begin{figure}[t]
\centering
\includegraphics[width=0.48\textwidth]{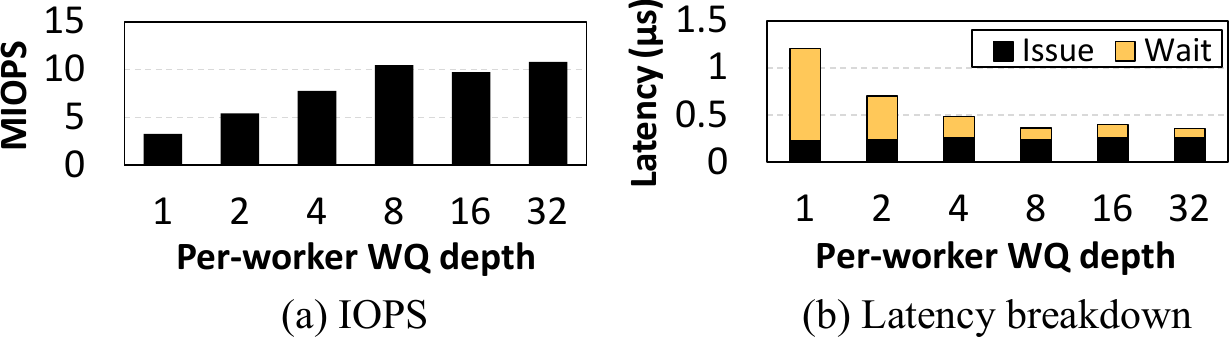}
\caption{Effect of \proposed{}’s asynchronous copy offloading on (a) worker throughput and (b) the resulting worker-side latency breakdown as the per-worker WQ depth varies.}
\label{fig:dsa_async_offloading}
\end{figure}

\begin{figure}[t]
\centering
\includegraphics[width=0.48\textwidth]{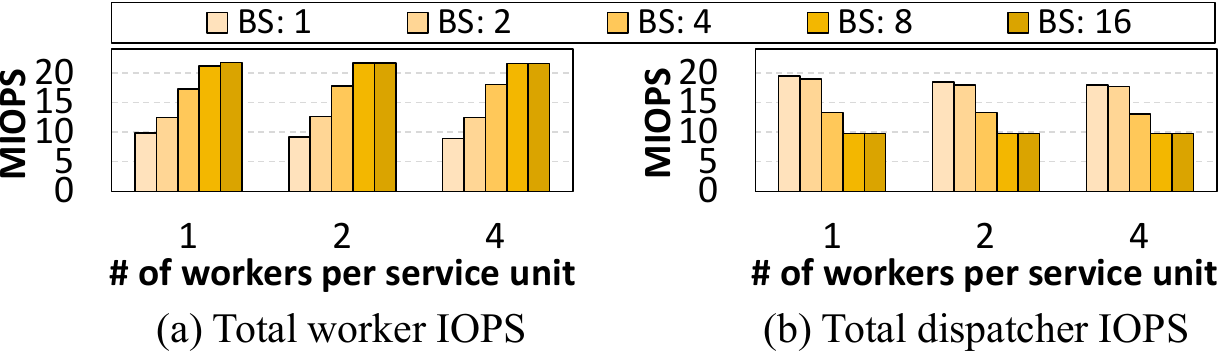}
\caption{Effect of DSA’s batch size (\texttt{BS}) on (a) total worker throughput and (b) total dispatcher throughput across four service units instantiated on a single DSA device.}
\label{fig:dsa_batched_offloading}
\vspace{-0.2em}
\end{figure}

{\bf Effect of asynchronous and batched offloading.} \fig{fig:dsa_async_offloading}(a) shows the benefits of \proposed{}’s asynchronous copy offloading by showing total worker IOPS as a function of per-worker WQ depth, assuming a DSA batch size of 1 and one worker per service unit. Compared to a per-worker WQ depth of one (effectively equivalent to synchronous offloading), total worker IOPS increases by 3.3$\times$ at a WQ depth of 32. This improvement is primarily driven by a significant reduction in wait time for a copy descriptor to complete (\texttt{Wait} in \fig{fig:dsa_async_offloading}(b), reduced by up to $9.8\times$ on average), enabled by overlapping multiple DSA-side data transfers, while copy-descriptor issue latency (\texttt{Issue}) remains nearly unchanged.

\fig{fig:dsa_batched_offloading}(a) further highlights our batch descriptor scheme's effectiveness. To maximize concurrency across configurations, we partition each DSA group’s maximum WQ depth (i.e., 32 descriptors) among the workers in the corresponding service unit while varying the number of workers per unit. We make two key observations. First, batching substantially improves offloading efficiency, saturating total worker IOPS at approximately 22.1 MIOPS. Second, a single worker thread (i.e., single CPU core) per service unit is sufficient to achieve peak performance by maximizing both per-worker WQ depth and batch size. These results confirm that our API enables high-throughput copy offloading with fewer CPU threads (cores) by hiding data transfer latency via asynchronous offloading and amortizing copy descriptor issue overhead through batching.

{\bf Dispatcher vs. worker interference.} Although our microbenchmark is configured to achieve 20 MIOPS of dispatcher throughput, the realized throughput gradually degrades as the batch size increases (\fig{fig:dsa_batched_offloading}(b), showing a 1.9$\times$ reduction when the worker batch size increases from 1 (non-batched) to 16). This degradation arises because the dispatcher and workers within a service unit share the same DSA engine; thus, aggressive worker offloading with large WQ depths and batch sizes can delay dispatcher progress. Consequently, the effective throughput across all service units is bounded by the lower of total dispatcher and worker throughput, reaching at most 13.3 MIOPS in practice. We therefore use 13.3 MIOPS as a per-DSA throughput target for configuring \proposed under GPU-initiated I/O; specifically, one worker per service unit with a batch size of around 4 is sufficient to saturate peak DSA performance and achieve this 13.3 MIOPS target.

\subsection{Aggregated Timing Model Updates}
\label{subsect:aggregated_timing_model_updates}

\proposed{}'s timing model builds on top of NVMeVirt’s simple timing model (\sect{subsect:nvmevirt}), parameterized by maximum throughput and minimum latency. However, \proposed{}'s distributed frontend architecture introduces additional system design considerations for properly updating the timing model across multiple dispatchers.

{\bf Global timing model.} One possible design point is to use a \emph{local} timing model per dispatcher, with each of the $N$ timing models configured to sustain $1/N$ of the target throughput. While this approach simplifies the design by eliminating inter-dispatcher dependencies, it cannot faithfully model skewed request distributions in which I/O requests are concentrated in only a subset of SQs and are therefore handled by only a few dispatchers. Even if each active SQ is heavily loaded enough for a real SSD to achieve high IOPS through sufficient intra-queue parallelism, the emulator with local timing models remains constrained by the aggregate capacity of only the few active timing models. Consequently,  we propose a \emph{global} timing model shared across dispatchers so that it can capture the global system load faithfully. With such design, the timing model’s internal state, such as SSD resource
availability, is shared across dispatchers and must therefore be updated consistently to avoid race conditions. To this end, we protect the shared state with a lock, serializing timing model updates across dispatchers on a per-request basis. However, this design introduces a tension between frontend scalability and serialization overhead: sustaining high frontend IOPS requires multiple dispatchers, but at high request rates, those dispatchers contend more frequently for the shared lock, increasing serialization overhead and threatening scalability.

{\bf Aggregated timing model updates.} To resolve this tension, we propose \emph{aggregated timing model updates}, which preserve scalability by mitigating serialization overhead while preserving timing model semantics. Combined with our dispatcher's coalesced request fetching feature, this design allows each dispatcher to enter the critical section of timing model update once per fetched set of requests rather than once per request, as in the original design. Specifically, after fetching a set of requests, a dispatcher first computes the corresponding state updates (i.e., aggregate scheduling-time increment for each scheduling instance in \fig{fig:nvmev_architecture}(b)), assuming back-to-back scheduling of requests on their target instances. It then acquires the lock and applies these updates in a single step, thereby amortizing serialization overhead. Finally, each request's completion time is determined by assuming that requests are assigned to scheduling instances in the order in which they appear in the SQ.

\section{Methodology}
\label{sect:methodology}

{\bf System configurations.} We conduct our experiments on a server with an Intel Xeon 6787P 86-core single-socket CPU and 256 GB of DDR5 DRAM. The CPU includes four DSA devices (v2.0), and we configure each DSA device into four DSA groups, each with one engine and two WQs. Each WQ is set to a maximum depth of 32 and operates in dedicated mode, as \proposed uses DSA exclusively within the kernel. The server contains an NVIDIA H200 GPU~\cite{h200} for GPU-initiated I/O and a 1.92 TB Solidigm D7-PS1010 PCIe Gen5 SSD~\cite{solidigm_d7}, which we use to validate the performance modeling fidelity of \proposed on a modern enterprise-grade SSD.

We reserve 33 CPU cores and 128 GB of DRAM for SSD emulation in both the baseline NVMeVirt and the proposed \proposed systems. In both designs, all dispatcher and worker threads are pinned to separate CPU cores. NVMeVirt uses one dispatcher and 32 workers, whereas \proposed uses up to 16 service units, each consisting of one dispatcher and one worker, with each service unit paired with a DSA group. In \proposed, a single worker per service unit, combined with asynchronous batched offloading, suffices to saturate DSA performance. By default, each worker uses an offloading batch size of 16, larger than the batch size of 4 used in \sect{subsect:dsa_accelerated_data_transfer}, because end-to-end emulation introduces additional worker-side tasks beyond data transfer, slightly reducing the batch descriptor issue rate compared to our microbenchmark. Each worker maintains a WQ depth of 32, while the dispatcher uses synchronous, non-batched offloading for request fetching.

{\bf Benchmarks.} We evaluate random storage read benchmarks using fio (v3.38)~\cite{fio} for CPU-centric I/O and BaM~\cite{bam} for GPU-initiated I/O. For compatibility with our Linux kernel version (v6.16.\allowbreak10), we use a patched BaM implementation based on commit \texttt{315fadf}. We use fio with a 4 KB I/O size and the SPDK engine (v25.09)~\cite{spdk}. In fio, we scale request-level parallelism by increasing the I/O depth while fixing the number of fio threads at 32. With SPDK, each fio thread uses a dedicated I/O queue pair, resulting in 32 queue pairs, each with a queue depth of 1,024. We use BaM with a default I/O size of 512 bytes, 256 I/O queue pairs, matching the maximum supported by the D7-PS1010 SSD, and a queue depth of 1,024. This configuration supports on the order of $100\text{K}$ concurrent requests. Accordingly, we set the default number of GPU threads to 256K.

\section{Evaluation}
\label{sect:evaluation}

In this section, we first validate \proposed{}’s ability to capture the performance characteristics of a modern PCIe Gen5 SSD (\sect{subsect:validation}), and then evaluate its scalability to next-generation IOPS levels (\sect{subsect:scalability}). Next, we quantify the contribution of each proposed optimization to \proposed{}'s achievable IOPS (\sect{subsect:ablation_study}). Finally, we present sensitivity studies to further characterize its performance across a range of I/O stack configurations (\sect{subsect:sensitivity_study}).

\subsection{Emulator Validation}
\label{subsect:validation}

We first validate \proposed against a real SSD by profiling the performance characteristics of the D7-PS1010 using the fio and BaM benchmarks. We then execute the same benchmarks on NVMeVirt and \proposed, with their timing models calibrated to reproduce the D7-PS1010’s characteristics, specifically a minimum latency of 50 $\mu s$ and a peak random-read throughput of 2.47 MIOPS.

\begin{figure}[t]
\centering
\includegraphics[width=0.48\textwidth]{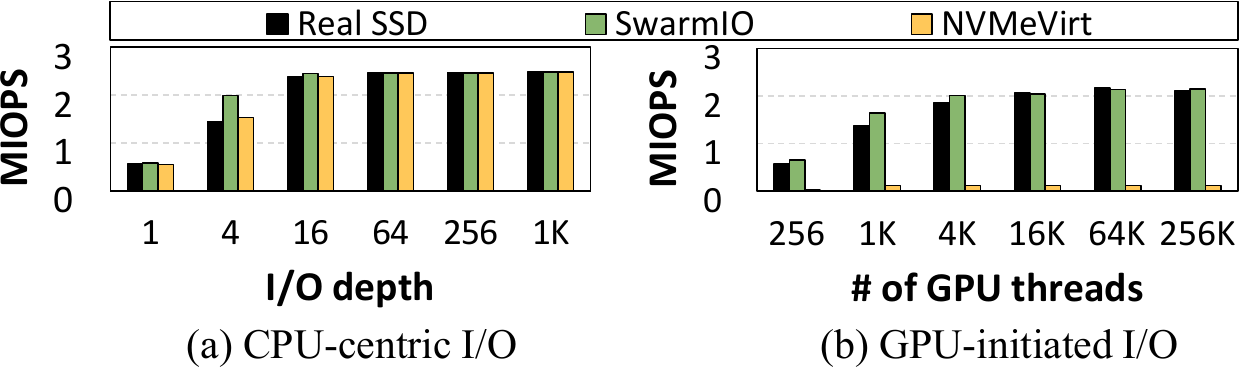}
\caption{Sustained IOPS under (a) CPU-centric I/O (fio) and (b) GPU-initiated I/O (BaM) for \proposed and NVMeVirt, compared with a real SSD (Solidigm D7-PS1010).}
\label{fig:swarmio_validation_iops}
\vspace{-0.3em}
\end{figure}

\begin{figure}[t]
\centering
\includegraphics[width=0.45\textwidth]{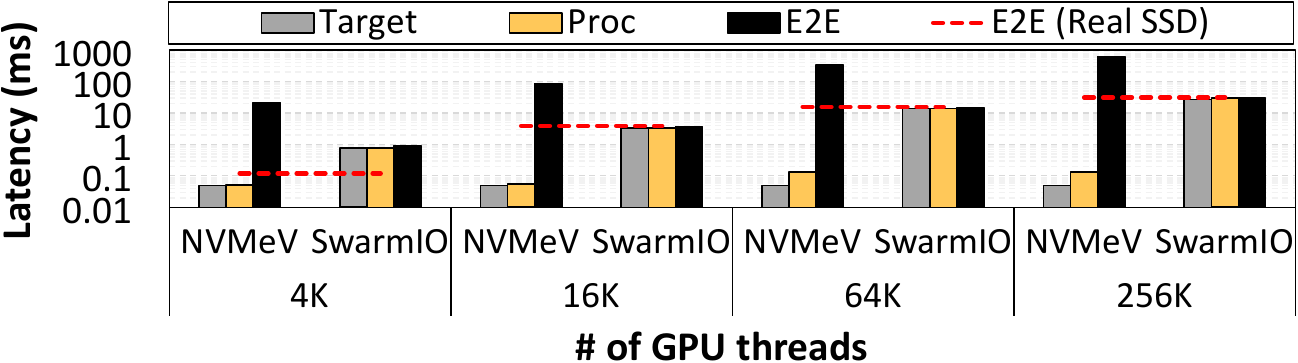}
\caption{Average target completion latency from the timing model (\texttt{Target}), request processing time to meet the target (\texttt{Proc}), and end-to-end latency (\texttt{E2E}) of NVMeVirt and \proposed under GPU-initiated I/O. The red dotted line denotes the end-to-end latency measured on the real SSD.}
\label{fig:swarmio_validation_breakdown}
\end{figure}

{\bf Throughput.} As shown in \fig{fig:swarmio_validation_iops}, \proposed effectively captures the IOPS characteristics of a commodity SSD under both CPU-centric I/O and GPU-initiated I/O. The average relative IOPS error is 7.7\% for fio and 7.4\% for BaM, indicating close alignment with the real SSD in both settings. Although \proposed exhibits slightly higher IOPS under light input loads (e.g., at I/O depths below 16 in fio and with fewer than 4K GPU threads in BaM), it closely matches the SSD IOPS under highly parallel conditions, which is typical of future IOPS-optimized SSD use cases. NVMeVirt is also sufficiently accurate to capture the SSD IOPS in CPU-centric I/O, with an average relative IOPS error of 1.2\% for fio. However, under GPU-initiated I/O, it fails to reach the configured target IOPS regardless of the degree of GPU thread-level parallelism, due to the bottlenecks discussed in \sect{subsect:challenges_with_bam}.

{\bf Latency.} NVMeVirt's limited frontend scalability causes requests to be fetched at a rate far below the submission rate under high load, thereby incurring substantial SQ queuing delays. GPU-initiated I/O further exacerbates this bottleneck. \fig{fig:swarmio_validation_breakdown} illustrates the average target completion latency (\texttt{Target}), internal request processing time used to realize \texttt{Target} (\texttt{Proc}), and end-to-end latency including SQ queuing delay (\texttt{E2E}) for NVMeVirt and \proposed under GPU-initiated I/O. Because the single dispatcher becomes the bottleneck in NVMeVirt, the timing model remains in the ``low-load'' mode shown in \fig{fig:nvmev_architecture}(b) regardless of the true system load, and thus always derives \texttt{Target} as the configured minimum latency of 50~$\mu$s. As the number of GPU threads increases, \texttt{Proc} gradually deviates from \texttt{Target}, reaching up to 2.7$\times$ higher, suggesting that NVMeVirt’s backend cannot sustain that degree of request-level parallelism and thus fails to meet the target completion time. \texttt{E2E} also greatly exceeds \texttt{Proc}, indicating that SQ queuing delay dominates the overall latency in NVMeVirt. Consequently, NVMeVirt’s \texttt{E2E} is on average 21.8$\times$ higher than that of the real SSD (\texttt{E2E (Real SSD)}), showing that SQ queuing delay severely undermines modeling fidelity. In contrast, \proposed prevents the dispatchers from becoming a bottleneck, allowing its timing model to derive target completion latencies that closely match with the target SSD’s average completion latency at each input load level. In addition, DSA-accelerated data transfers allow the workers to satisfy these target latencies, keeping \proposed{}'s end-to-end latency remains close to that of the real SSD, with an average relative error of 2.8\%.

\subsection{Scalability}
\label{subsect:scalability}

\begin{figure}[t]
\centering
\includegraphics[width=0.42\textwidth]{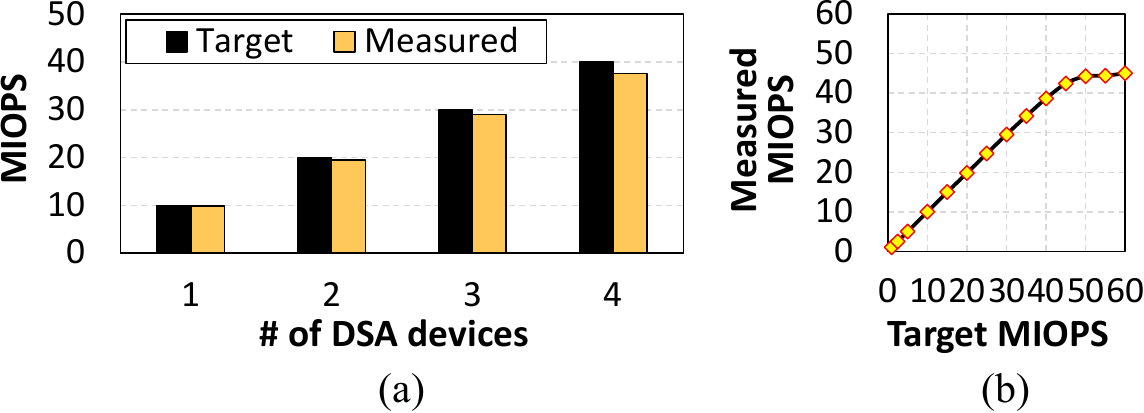}
\caption{\proposed{}'s (a) peak IOPS as the number of DSA devices increases, and (b) sustained IOPS with four DSA devices when the target IOPS is scaled in 5 MIOPS increments.}
\label{fig:swarmio_scalability}
\vspace{-0.3em}
\end{figure}

We next evaluate the scalability of \proposed. As shown in \fig{fig:swarmio_scalability}(a), under GPU-initiated I/O, \proposed achieves up to 9.8 MIOPS with a single DSA device, compared to the 10 MIOPS target. This is close to the practical peak DSA performance under dispatcher--worker interference reported in \sect{subsect:dsa_accelerated_data_transfer}, i.e., the minimum of dispatcher and worker throughput when both operate concurrently. To scale beyond a single DSA device, we increase the number of DSA devices and, accordingly, the total number of service units. Specifically, each DSA device serves four service units, each mapped to a dedicated DSA group. With four DSA devices (the maximum available in our server), \proposed achieves up to 38.6 MIOPS at a 40 MIOPS target, corresponding to a 303.9$\times$ speedup over NVMeVirt. It is worth pointing out that \proposed is not limited to emulating only the maximum performance supported by the DSA devices. Using four DSA devices, it can flexibly scale the target performance to emulate storage across a range of operating points, as shown in \fig{fig:swarmio_scalability}(b), where we increase the target IOPS of the timing model in 5 MIOPS increments starting from 5 MIOPS. \proposed sustains more than 96.6\% of the configured target IOPS up to 40 MIOPS. Beyond this point, it can still reach up to 45 MIOPS but no longer consistently matches the configured target. While \proposed is currently limited to 40 MIOPS, this limit is primarily imposed by the available DSA resources on our evaluation platform, rather than by an architectural limitation of \proposed. Given its scalable design, we expect \proposed to reach 100 MIOPS targets under a scale-up configuration, assuming a dual-socket platform provisioned with at least five DSA devices per socket.

\subsection{Ablation Study}
\label{subsect:ablation_study}

We conduct ablation studies to quantify the impact of each proposed optimization on frontend scalability and to demonstrate the effectiveness of aggregated timing model updates.

\begin{figure}[t]
\centering
\includegraphics[width=0.45\textwidth]{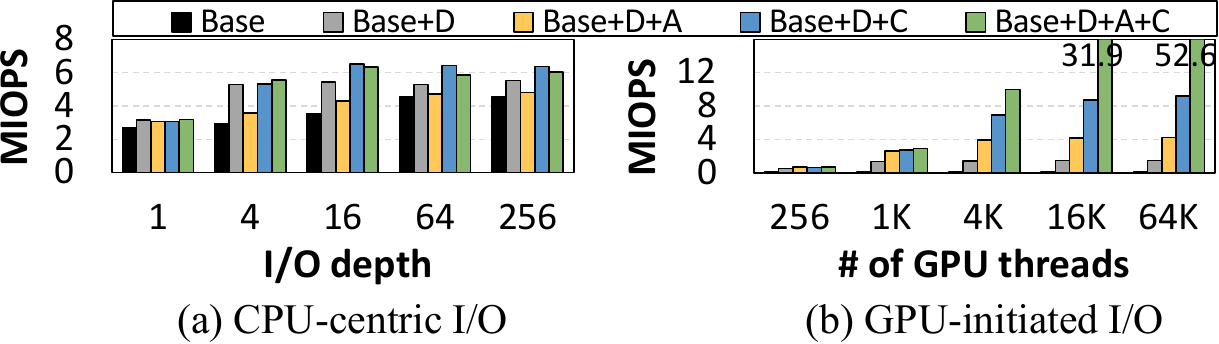}
\caption{Effect of \proposed{}’s optimizations on baseline NVMeVirt frontend performance (\texttt{Base}) by adding (\texttt{D}) a distributed architecture with 16 dispatchers, (\texttt{A}) DSA-accelerated request fetching, and (\texttt{C}) coalesced request fetching.}
\label{fig:swarmio_ablation_frontend}
\end{figure}

{\bf Frontend performance.} \fig{fig:swarmio_ablation_frontend} shows how frontend throughput, excluding backend data transfers, improves as we add the following features to the baseline NVMeVirt (\texttt{Base}): (\texttt{D}) a distributed architecture with 16 dispatchers, (\texttt{A}) DSA-accelerated request fetching with synchronous offloading, and (\texttt{C}) coalesced request fetching that allows up to 1,024 requests within an SQ to be fetched at once. Our analysis shows that, under CPU-centric I/O, the distributed design with coalesced fetching (\texttt{D+C}) alone, without DSA-accelerated request fetching, is sufficient to saturate the throughput supported by 32 CPU threads, reaching up to 6.5 MIOPS at an I/O depth of 16 or higher. Under GPU-initiated I/O, the distributed design alone does not scale beyond 1.5 MIOPS. Adding either DSA-accelerated request fetching (\texttt{D+A}) or coalesced fetching (\texttt{D+C}) to the distributed-only design improves frontend scalability, delivering speedups of up to 2.8$\times$ and 6.2$\times$, respectively. Importantly, combining both optimizations with the distributed design (\texttt{D+A+C}) increases frontend throughput to at most 52.6 MIOPS, corresponding to a 537.2$ \times$ speedup over the baseline. This speedup is particularly pronounced because large PCIe P2P DMA operations improve transfer efficiency, while our high-throughput offloading API and DSA's pipelined processing effectively hide data transfer latency. Together, these results show that the proposed techniques work synergistically to improve frontend performance.

\begin{figure}[t]
\centering
\includegraphics[width=0.44\textwidth]{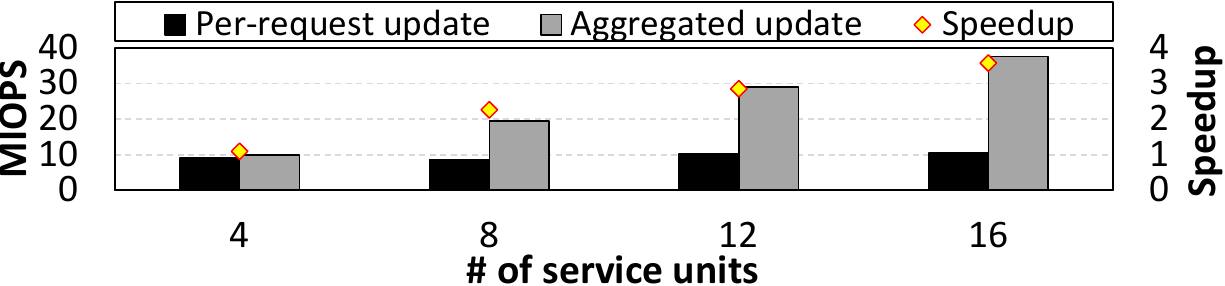}
\caption{Effect of \proposed{}'s aggregated timing model updates relative to the baseline per-request updates on achievable IOPS under GPU-initiated I/O, while varying the number of service units sharing a global timing model.}
\label{fig:swarmio_ablation_timing_model}
\end{figure}

{\bf Aggregated timing model updates.} In addition, we evaluate the effectiveness of our aggregated timing model updates.  In \fig{fig:swarmio_ablation_timing_model}, we scale the number of service units, along with the corresponding number of DSA groups, and set the target IOPS accordingly, i.e., 10 MIOPS for every four service units. We then measure the achievable IOPS for designs with and without aggregated timing model updates. Without aggregated updates, per-request timing model updates incur increasing contention as the number of dispatchers grows, preventing performance from scaling beyond 10 MIOPS. In contrast, aggregated timing model updates substantially reduce serialization overhead across dispatchers. With 16 service units, \proposed scales to sustain the target of 40 MIOPS, achieving a 3.6$\times$ speedup over the baseline design.

\subsection{Sensitivity Study}
\label{subsect:sensitivity_study}

\begin{figure}[t]
\centering
\includegraphics[width=0.45\textwidth]{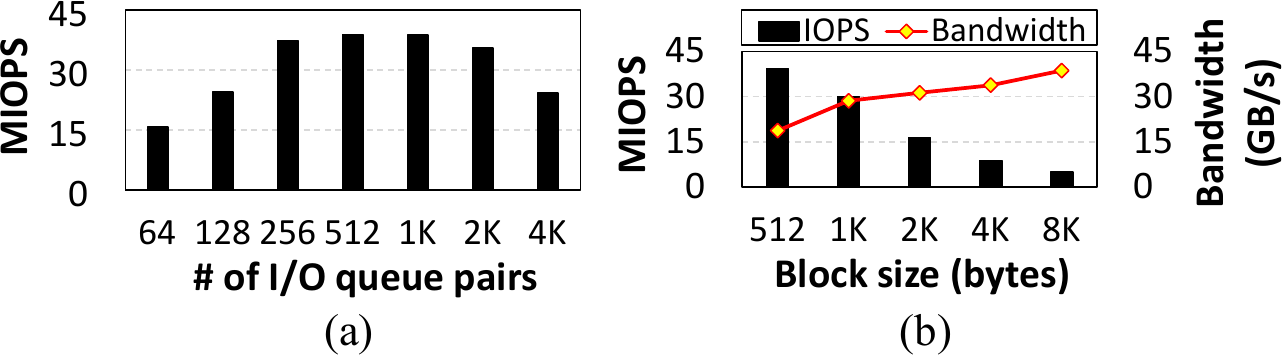}
\caption{Sensitivity of \proposed{}’s sustained IOPS to (a) the number of I/O queue pairs and (b) I/O block size under GPU-initiated I/O with 2M threads.}
\label{fig:swarmio_sensitivity}
\vspace{-0.2em}
\end{figure}

We further examine the robustness and practical limits of \proposed under different I/O system configurations. Specifically, we study the impact of the number of I/O queues and the SSD block size (i.e., I/O granularity). The former is important because future IOPS-optimized SSDs may require a wider NVMe host interface to fully exploit internal parallelism, while the latter matters because the optimal I/O size can vary across applications. We use the BaM benchmark because GPU-initiated I/O naturally exercises both dimensions by scaling the number of GPU threads to utilize more I/O queues and by varying the I/O size via the GPU-side I/O buffer size.

{\bf Number of I/O queues.} \fig{fig:swarmio_sensitivity}(a) shows the robustness of \proposed{}’s performance modeling across various I/O queue counts, with requests submitted by 2M GPU threads. \proposed sustains near-peak performance up to 1K queues, enabled by our throughput-optimized frontend architecture combining distributed dispatching with coalesced request fetching via DSA acceleration. Given an I/O depth of 1K, this is sufficient for 1M GPU threads to enqueue requests simultaneously. Beyond 1K queues, achievable IOPS falls below the target (e.g., reaching only 88.9\% of the 40 MIOPS target at 2K queues) because the sequential overhead of processing many SQs begins to constrain frontend dispatch throughput.

{\bf Block size.} \fig{fig:swarmio_sensitivity}(b) shows overall performance as the block size varies. As the block size increases, achievable IOPS drops from the peak at 512 bytes. For example, \proposed achieves 15.1 MIOPS with 2 KB blocks. We attribute this limit to practical platform constraints when DSA devices transfer data to a peer PCIe device (i.e., GPU) across distinct PCIe controllers to which the devices are attached. As the block size increases, transfer efficiency improves, and the achievable bandwidth correspondingly rises, but only up to about 40 GB/s. Specifically, through stress testing with four DSA devices, we find that the aggregate DSA bandwidth for copying data to a PCIe Gen 5 $\times$16 GPU reaches up to 42.1 GB/s at I/O sizes of 8 KB and above, still well below the GPU's theoretical unidirectional PCIe bandwidth of 64 GB/s. Nevertheless, \proposed achieves at least 44.6\% of the theoretical peak bandwidth for I/O sizes greater than 1 KB, a level that no modern enterprise-grade SSD provides at such I/O sizes. Moreover, according to the StorageNext roadmap~\cite{storagenext,fms2025_nvidia,ocp2025_nvidia}, future IOPS-optimized SSDs are expected to target ultra-high IOPS at 512-byte granularity, and our emulator sustains 38.6 MIOPS at this operating point.

\section{Case Study: Scaling Up GPU-centric Storage Systems with Future IOPS-optimized SSDs}
\label{sect:case_study}

As introduced in \sect{sect:introduction}, GPU-initiated I/O is a promising mechanism for data-intensive applications with highly parallel, sparse, and random access patterns. To understand the end-to-end benefits of combining next generation, ultra-high IOPS-optimized SSDs with GPU-initiated I/O in such applications, we conduct a case study on GPU-accelerated, on-disk vector search~\cite{diskann,starling,fusion_anns,turbocharge_vectordb,pipeann}.

{\bf GPU-accelerated, on-disk vector search.} As a representative case study, we focus on vector search, a core component of RAG systems for retrieving data (e.g., documents or images, that is semantically relevant to a query in the embedding space). We utilize CAGRA~\cite{cagra}, a GPU-accelerated vector search algorithm based on graph-based approximate nearest neighbor search (ANNS). In graph-based ANNS, dataset vectors are indexed as a graph, and the CAGRA search traverses the graph toward nodes closer to the query to identify the top-$k$ nearest vectors. Upon visiting a node corresponding to a dataset vector, the algorithm (1) computes the distances between the query and the node’s neighbors, (2) sorts the results by distance, and (3) iteratively extends the search by proceeding to the closest, unvisited nodes until the configured maximum number of iterations is reached.

We consider an on-disk vector search scenario~\cite{diskann,starling,fusion_anns,turbocharge_vectordb,pipeann} in which the entire vector index (i.e., both the graph structure and vector data) cannot fit in GPU or CPU memory. To enable direct SSD access from the GPU, we integrate GPU-initiated I/O with CAGRA so that the GPU can read index data directly from the SSD. We use the BIGANN-100M dataset~\cite{bigann}, whose CAGRA index occupies 71.5 GB. To emulate a larger-scale deployment, we evaluate a downscaled configuration in which the GPU memory available for CAGRA search is limited to 2 GB. This preserves approximately the same GPU memory-to-index ratio as a 144 GB H200 GPU serving a 5 TB index, a practical scale for vector search over billions of vectors~\cite{hermes}.

\begin{figure}[t]
\centering
\includegraphics[width=0.42\textwidth]{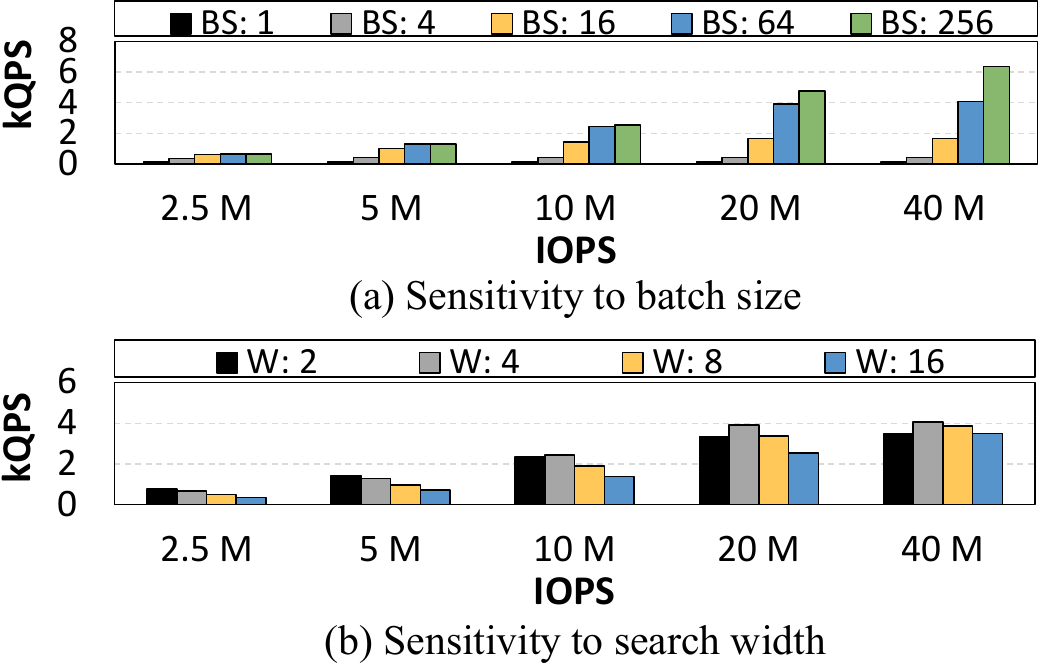}
\caption{End-to-end performance of on-disk CAGRA search when scaling \proposed target IOPS from 2.5 to 40 MIOPS: (a) sensitivity to batch size (\texttt{BS}) at search width 4, and (b) sensitivity to search width (\texttt{W}) at batch size 64.}
\label{fig:cagra_bam_performance}
\vspace{-0.5em}
\end{figure}

{\bf Implication of future IOPS-optimized SSDs.} We measure CAGRA search throughput, i.e., queries per second (QPS), while varying SSD IOPS using \proposed. In this experiment, we set the block size to 512 bytes, corresponding to the 128-dimensional FP32 vector size. As shown in \fig{fig:cagra_bam_performance}(a), increasing SSD IOPS provides little benefit at a small batch size (\texttt{BS}) of 4 because the CAGRA search generates insufficient parallel storage I/O to utilize the additional IOPS. As batch size increases, however, the benefit of higher SSD IOPS becomes clear. Scaling SSD IOPS by 16$\times$, from 2.5 to 40 MIOPS, yields a 9.7$\times$ end-to-end speedup of CAGRA search at a batch size of 256. In other words, higher storage IOPS enables larger batch sizes by allowing more vector fetches from storage to proceed in parallel, thereby improving QPS.

{\bf Looking forward: the need for IOPS-aware algorithmic shifts.} Increasing search width through \emph{beam search}~\cite{diskann,cagra}, which expands the search frontier by visiting multiple graph nodes in parallel at each iteration, allows the algorithm to explore more nodes per iteration and can potentially reduce the total number of search iterations needed to reach a target accuracy. In ANNS, recall is the primary accuracy metric and is defined as the fraction of ground-truth top-$k$ neighbors included in the retrieved results. We first profile CAGRA search offline by varying the search width and identifying, for each search width, the minimum number of search iterations required to guarantee at least 95\% recall. We then apply the corresponding iteration count for each search width and measure QPS while varying both search width and SSD IOPS. As shown in \fig{fig:cagra_bam_performance}(b), the optimal search width (\texttt{W}) depends on the provisioned IOPS. For example, a search width of 2 is optimal at 5 MIOPS or below, whereas a search width of 4 becomes faster at 10 MIOPS or above. This result suggests that future ultra-high IOPS SSDs can shift the optimal algorithmic configuration for maximizing end-to-end performance, highlighting the need for IOPS-aware algorithmic shifts in data-intensive applications running on next generation GPU-centric storage systems.

\section{Related Work}
\label{sect:conclusion}

There exists a large body of prior work on SSD performance modeling, and numerous prior studies have also explored GPU-centric storage systems, which represent the primary target environments for \proposed. We briefly summarize the most relevant studies below.

{\bf SSD performance modeling.}
Prior work on SSD performance modeling spans both simulation and emulation. Several studies~\cite{mqsim,simplessd, nandflashsim, wiscsim, ssdmodel, ssdsim, flashsim} develop detailed SSD simulators for modeling SSD's internal device-level behavior, while integrating them into full-system simulation environments. While effective for modeling internal SSD behavior, they do not support real-time modeling of GPU-centric storage systems. Prior emulators~\cite{flexdrive,vssim,femu,nvmevirt} instead expose software-defined SSDs to systems for end-to-end evaluation. FEMU~\cite{femu} adopts a QEMU-based virtualization approach, while NVMeVirt~\cite{nvmevirt} presents a software-defined PCI NVMe device to the host system and supports versatile storage emulation environments. Recent studies further extend SSD emulation to emerging interfaces and device classes~\cite{znsplus,confzns,cemu,cyclon}, but still primarily target conventional CPU-centric storage settings rather than GPU-centric storage systems. Some prior works~\cite{fssd, openexpress, openssd} utilize FPGA-based platforms for SSD prototyping. While they offer hardware acceleration with native PCIe connectivity, they require specialized hardware and expertise. A growing body of work uses these modeling frameworks to study real-world applications whose performance depends on SSD designs not yet commercially available~\cite{instattention, ecssd, megis, graphssd, assasin}. \proposed broadens this line of work by enabling performance modeling for a new class of applications with next generation ultra-high IOPS demands in GPU-centric storage environments.

{\bf GPU-centric storage systems.}
Numerous prior studies explore GPU-centric storage systems to meet the high random-access IOPS demands of data-intensive workloads. BaM~\cite{bam} enables high-IOPS GPU-initiated I/O for fine-grained random accesses. GIDS~\cite{gids} applies GPU-initiated I/O to GNN training with sparse and irregular accesses, and GMT~\cite{gmt} proposes a GPU-orchestrated memory hierarchy spanning GPU memory, host memory, and SSDs. Other studies propose GPU-centric file systems~\cite{geminifs,gofs} or extend GPU-initiated I/O with asynchronous execution to overlap I/O and computation~\cite{cam,agio}. There is also a growing industry trend toward satisfying the IOPS demand of such systems at the storage-device level~\cite{storagenext}. Our work stands apart from this literature by focusing on end-to-end performance modeling of GPU-centric storage systems while enabling flexible scaling of storage IOPS.

\section{Conclusion}
\label{sect:conclusion}

GPU-initiated I/O introduces a fundamentally different operating regime for storage systems, where massive numbers of GPU threads generate fine-grained requests at extreme rates. Existing SSD emulation frameworks are not designed to handle this level of parallelism and IOPS, limiting their effectiveness in studying emerging GPU-centric workloads. We present \proposed, an SSD emulation framework that sustains tens of millions of IOPS while preserving accurate end-to-end behavior. Our evaluation shows that it closely matches real SSD performance and enables scalable exploration of GPU-centric storage. \proposed provides a practical foundation for studying IOPS-intensive workloads and highlights the growing importance of storage scalability in GPU-centric computing.

\section*{Acknowledgment}
This work was supported in part by SK hynix, which provided funding for the study and design of \proposed, and in part by Institute of Information \& Communications Technology Planning \& Evaluation(IITP) grant funded by the Korea government(MSIT) (No.RS-2024-00438851, (SW Starlab) High-performance Privacy-preserving Machine Learning System and System Software), (No.RS-2025-02264029, Implementation and Validation of an AI Semiconductor-Based Data Center Composable Cluster Infrastructure, 30\%), which provided the computing infrastructure (CPU and GPU) used in this study.

\bibliographystyle{IEEEtranS}
\bibliography{refs}

\end{document}